\documentclass[number, 3p]{elsarticle}
\usepackage[english]{babel}
\usepackage[utf8]{inputenc}
\usepackage{amsmath}
\usepackage{amssymb}
\usepackage{graphicx}
\usepackage{url}
\usepackage[normalem]{ulem}
\usepackage{soul}

\usepackage[usenames,dvipsnames,svgnames,table]{xcolor}

\usepackage{color}

\renewcommand{\epsilon}{\varepsilon}

\begin{document}
\title{Effectiveness of wealth-based {\em vs} exchange-based tax systems in reducing inequality}
\author[utfpr]{Thiago Dias}
\ead{thiagodias@utfpr.edu.br}
\author[ufrgs]{Sebastián Gonçalves}
\ead{sgonc@if.ufrgs.br}

\address[utfpr]{Universidade Tecnológica Federal do Paraná, Campus Dois Vizinhos, Estrada para Boa Esperança, km 04, 85660-000, Dois Vizinhos PR, Brazil}
\address[ufrgs]{Instituto de Física, Universidade Federal do Rio Grande do Sul, Caixa Postal 15051, 91501-970 Porto Alegre RS, Brazil}
\date{\today}
\cortext[ca]{Corresponding author}

\begin{abstract}
In the so-called ``fair'' models of peer-to-peer wealth exchanges, economic inequality  tends to reach its maximum value asymptotically.  This global trend is evident as the richest continuously accumulate a larger share of wealth at the expense of others. To address the mounting issue of inequality, different strategies of taxes and redistribution are commonly employed. 
Our study delves into the interplay between wealth and trade (consumption) tax bases, probing their impact on wealth distribution within wealth-conservative economies. The ultimate aim is to unearth an optimal framework that adeptly curbs inequality.
Through a meticulous analysis of varying tax rates and the allocation of the collected tax to the most economically vulnerable strata, we unveil a compelling pattern resembling two distinct phases.
These phases delineate the most effective systems for inequality mitigation. Our findings underscore the synergistic potential of amalgamating these tax systems, surpassing the individual efficacy of each. This synthesis beckons policymakers to weave together tax rates and precision-targeted redistribution, crafting tax systems that wield the potential for tangible and substantial reductions in economic disparity.
\end{abstract}


\maketitle

\section{Introduction}

Empirical studies on the distribution of wealth of individuals in different economies demonstrate the presence of two classes. While the wealthiest 1\% to 5\% of the population follows a power-law distribution, the distribution of the remaining fits in a Gamma distribution \cite{Alvaredo2018, capitalXXI, saez2016wealth,WorldBank}. The increasing gap between the wealth held by each class is a worldwide concern. For instance, in 2020, 43\% of the world's wealth was in the hands of the richest 1\% of the population. Two years later, it increased to roughly 47\% \cite{credit-suisse20,credit-suisse22}. Would this trend continue indefinitely, a small group of persons would possess all the economic resources while the rest nothing. 

A common measure of inequality is the Gini coefficient, $G$, which takes into account the relative mean absolute difference of income or wealth of individuals or households. It can assume values from 0 (perfect equality) to 1, where only one individual retains all the available wealth. When $G = 1$, the system is said to be in the condensate state. The prospect of condensation is profoundly undesirable since it implies a situation where a large majority of the population becomes trapped in poverty and the wealth exchange is severely reduced. Such an scenario represents the economic collapse \cite{iglesias-epjb-2012, scafetta-arxiv-2022}.

\begin{figure}[!htbp]
    \centering
    \includegraphics{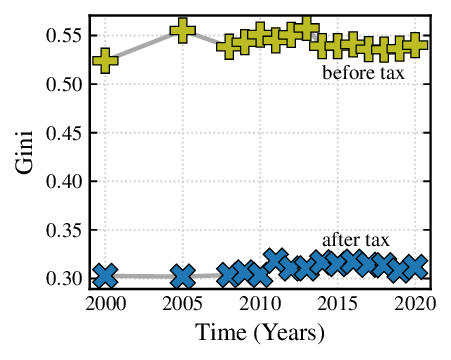}
    \caption{Evolution of the mean Gini coefficient of countries between 2000 and 2020, before and after taxation and redistribution, from OECD data \cite{oecd-stats}.}
    \label{fig1_gini}
\end{figure}

One approach that countries have adopted to combat the growth of inequality is the implementation of taxes and subsequent redistribution. Figure~\ref{fig1_gini} presents the evolution of the mean Gini coefficient of the OECD (The Organization for Economic Cooperation and Development) countries from 2000 to 2020 before and after taxation and redistribution. Data was taken from reference~\cite{oecd-stats}. It demonstrates that transfers of taxes lead to a reduction in the Gini coefficients by approximately 40\%. This result highlights the importance of implementing effective taxation and transfer policies in the pursuit of a more equitable society.

Various taxation systems can be observed globally, each with its own characteristics and implications \cite{cantante-hssc-2020, dianov-sus-2022}. The history of the discussion regarding the efficiency and fairness of different taxation policies dates back to 1776, when Adam Smith published his book \emph{The Wealth of Nations} \cite{smith-1776}. It remains a subject of discussion through the centuries to this day \cite{piketty-nber-2016, oliveira-epl-2017, guvenen-qje-2023, dammerer-cje-2023}. For example, Bradford and Toder \cite{bradford-nta-1976} analyzed the equity and simplicity of consumption and capital income-based taxes, arguing that consumption tax does not hinder capital formation and savings due to its neutrality between current and future consumption. Other studies have found that consumption-based taxation is more re-distributive than income-based taxation, including labor income \cite{bankman-slr-2006, bachas-prwp-2020}. The distinctions between wealth and capital income taxation were explored by Guvenen \emph{et al.}, who concluded that collecting taxes from wealth is preferable to taxing capital income since it can enhance productivity, promote economic growth, and lead to re-distributional gains. For the sake of simplicity, here we analyze the implications of exchange-based (equivalent to consumption-based) and wealth-based taxation systems. 

\begin{figure}[!htbp]
    \centering
    \includegraphics{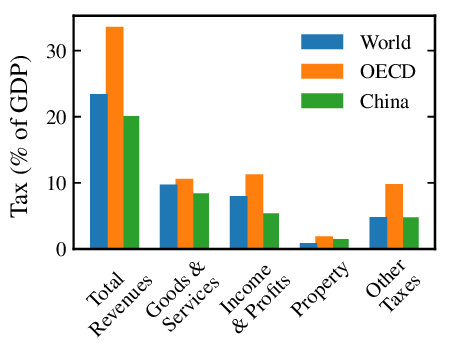}
    \caption{Tax collection and different taxation systems in 2020. Data is from reference~\cite{revenues-2020}.}
    \label{fig2_tax}
\end{figure}

Figure~\ref{fig2_tax} shows the percentages of taxes collected in 2020\footnote{The year 2020 was specifically chosen because it is the most recent and complete dataset accessible at the time we collected the data.}, including the data of the World, OECD countries, and China\cite{revenues-2020}. The figure presents the relative contribution of various tax categories to the total tax revenues. It reveals that property-related taxes represent a lower share compared to the other forms. Conversely, it can be observed that taxes on goods and services, as well as on incomes and profits are much higher and constitute the majority of the total taxes. The figure makes reference to other taxes, which are payroll, capital gain, and inheritance taxes, along with social security contributions. 

Econophysics has proven to be very valuable to model economic systems using the tools provided by Statistical Physics. Through first principles and numerical simulations, empirical data have been successfully reproduced. Moreover, researches on this field showed that unregulated markets inevitably evolve to condensation \cite{moukarzel-epj-2007, iglesias-epjb-2012, cardoso-csf-2023}. Most of the models consider money, properties, and capital (hereafter referred to as wealth, $w$) as an extensible variable related to agents and that can flow among them \cite{chakraborti-qf-2011}. Recent studies have shown an anti-correlation between inequality and liquidity, the average wealth exchanged per unit of time. In general, the liquidity tends to zero when $G \rightarrow 1$ \cite{bertotti-ejp-2016, biondi-jeic-2019, nener-ptrsa-2022}. Hence, the implementation of regulations to prevent condensation becomes crucial to maintain the dynamics of the economy \cite{scafetta-arxiv-2022}. 

This work primarily investigates the impact of wealth and exchanges taxation systems on inequality through agent-based simulations. We explore a combination of both to determine the most effective approach in reducing inequality within wealth-conservative systems. Before presenting these details, the model is explained in the next section.

\section{Kinetic exchange model with mixed taxation forms}\label{sec:model}

Our model consists in an ensemble of $N$ interacting agents with wealth $\{w_i\}$, $i = 1, 2, \ldots, N$. Agents are also characterized by their risk-aversion factor $\{\beta_i\}$, which sets the portion of wealth they are willing to put at stake during an exchange. At time $t = 0$, each agent is assigned a randomly distributed wealth and risk-aversion factor. While $\beta_i$ are in the interval $[0,1)$ and remains the same, $w_i$ must satisfy  $\sum_i^N w_i = 1$ and is exchanged throughout the simulation. The minimum wealth required for an agent to engage a trade is defined as $w_\text{min} = 1\times10^{-9}$. If an agent possesses less than $w_\text{min}$, its wealth is set to zero.

A Monte-Carlo step (MCS) is the unit of time when three processes take place: 
\begin{itemize}
    \item \emph{exchange of wealth}: pairs of agents, each with $w_i > w_\text{min}$, are randomly chosen and exchange part of their wealth;
    \item  \emph{tax collection}: a fraction is taken from the agents' wealth and/or from the amount traded during the exchange
    \item \emph{redistribution}: the tax collected is equally redistributed to the fraction $\tau$ of the poorest agents, also referred as target.
\end{itemize}

The yard-sale rule is employed to determine the amount exchanged between two agents, $i$ and $j$, ensuring a fair and equal opportunity for both parties \cite{hayes, cardoso-pa-2020}. This rule defines the amount traded ($\Delta w$) as
\begin{equation}
    \Delta w = \min[(1 - \beta_i)w_i, (1 - \beta_j)w_j].
    \label{delta}
\end{equation}
Each trade is taxed a fraction $\epsilon\lambda$ (tax on exchanges), where $\lambda$ is the total tax rate and $\epsilon = [0,1]$ is a variable that accounts for the taxation system.
With $\epsilon = 1$, taxes are collected solely on the exchanges. 
For $\epsilon = 0$, the taxes are applied just on the agent's wealth. 
If $0 < \epsilon < 1$, it is used a combination of the two systems.
 Considering two agents, $i$ and $j$, and assuming the former wins the exchange, the after-trade wealth's are
\begin{align}
    w_i^* &= w_i + (1 - \epsilon\lambda)\Delta w\nonumber\\
    w_j^* &= w_j - \Delta w,\label{exchange}
\end{align}
where $w_{i(j)}^*$ represents the wealth of the agent $i(j)$ after, and $w_{i(j)}$ before the transaction. It is worth noting that although it may appear that the winner pays the tax, this is not entirely accurate, as the tax is typically included in the price of the goods or services being exchanged. After the trade, the fraction $(1 - \epsilon)\lambda$ is taken from the wealth of all the agents. Finally, the amount collected as taxes is distributed equally among the $\tau N$ poorest agents.

The inequality is measured through the Gini coefficient \cite{sen-res-1976},
\begin{equation}
    G_k(t) = \frac{1}{2N\sum_i w_i(t)}\sum_{ij}^{N} | w_i(t) - w_j(t)|.
    \label{gini}
\end{equation}

The index $k$ in the Eq.~\ref{gini} indicates the taxation system in use. Specifically, $G_0$ represents a wealth-based tax system, and $G_1$ refers to a system where taxes are imposed only on trades or transactions. When no index is given, the Gini coefficient is calculated by considering a combination system of wealth and transaction.

The simulations were carried out with $N = 10^4$ interacting agents until equilibrium is achieved. We define equilibrium as the situation in which the mean time-step difference of $G$ is less than $1\times10^{-8}$ for $10^3$ MCS. The results are the average of $10^3$ independent samples for each set of the parameters ($\lambda$, $\tau$, and $\epsilon$).

\section{Taxes and redistribution effect on inequality}\label{sec:conservative}

In this section, we present the results of the above described model with the different tax systems.
First, for the wealth-based system, then for the transaction based system, and finally for the mixed system.
We emphasize that wealth is conserved, so no wealth is created or destroyed during the simulations.

\subsection{Wealth-based taxes}
\begin{figure}[!htbp]
    \centering
    \includegraphics{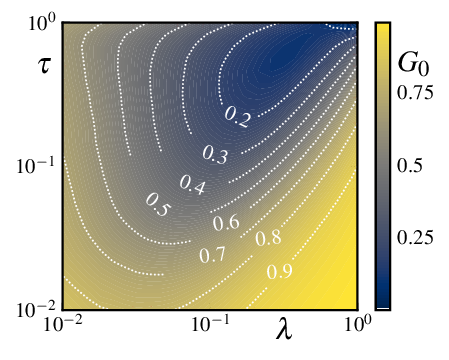}
    \caption{Contour plot of equilibrium $G_0$ (wealth tax base) as functions of the tax rate $\lambda$ and the fraction $\tau$ of poorest agents in log-log scale.}
    \label{fig:contourG0}
\end{figure}

Models with taxes on wealth have already been studied in different works. Specifically, we mention references~\cite{nener-ptrsa-2022} and~\cite{cardoso-pa-2020}. For comparative purposes, those results are reproduced here with our model. In Fig.~\ref{fig:contourG0}, we show the dependence of the Gini coefficient on $\lambda$ and $\tau$. It can be observed that for a significant reduction in $G$, both the tax rate and the fraction of agents participating in the re-distributive process should be higher than 4---5\%. 

Furthermore, Fig.~\ref{fig:contourG0} reveals a non-trivial relationship between $G_0$, $\lambda$, and $\tau$. That is, for each collected tax rate $\lambda$, there is an optimal target fraction $\tau$ which produces the lowest Gini coefficient $G_0$. Therefore, it is possible to find the line along which the variation of $G_0(\lambda,\tau)$ is the steepest (we show that curve in Fig.\ref{fig:targeted}(b)).

\subsection{Exchange-based taxes}
\begin{figure}[!htbp]
    \centering
    \includegraphics{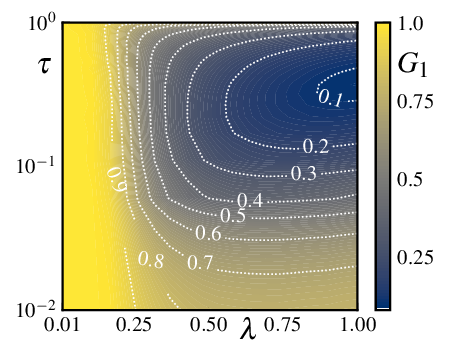}
    \caption{Contour plot of equilibrium $G_1$ (consumption tax base) as functions of the tax rate $\lambda$ and the fraction $\tau$ of poorest agents in log-linear scale.}
    \label{fig:contourGt}
\end{figure}

We present here the analysis of the effects of trade taxation ($\epsilon = 1$) on economic inequality measured by the Gini index $G_1$,  displayed in Fig.~\ref{fig:contourGt} as a function of $\lambda$ and $\tau$. Clearly, this taxation policy is inefficient when $\lambda < 0.25$, as $G_1$ exceeds 0.9 for almost every $\tau$ ranging from 0.01 to 1. Only when $\lambda \gtrapprox 1/3$ one can see values of $G_1 \leq 0.5$. It is important to emphasize that even a Gini value of 0.5 indicates a considerable level of economic inequality.
 
Clearly, as depicted in Fig.~\ref{fig:contourGt}, this tax system also possesses an optimal target fraction corresponding to the point of least inequality (smallest Gini) for a given tax rate. However, in somehow contrast to the corresponding plot of the wealth-base system (Fig.~\ref{fig:contourGt}), the optimal line of $G_1$ follows an almost horizontal path in the $\lambda$--$\tau$ space, which means that there is an almost constant target fraction, independent of the tax rate.

\begin{figure}[!htbp]
    \centering
    \includegraphics[width = 0.9\textwidth]{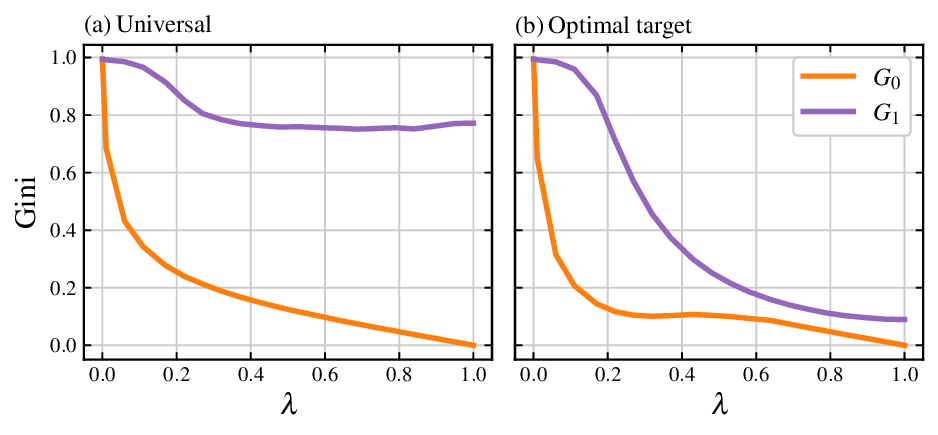}
    \caption{Equilibrium Gini coefficients for (a) universal and (b) optimal targeted redistributions considering taxation on wealth ($G_0$) and on trade ($G_1$).}
    \label{fig:targeted}
\end{figure}

Figure~\ref{fig:targeted} shows the impacts of those tax policies on the Gini coefficient with (a) universal redistribution, $\tau = 1$; and (b) optimal targeted redistribution. It is noteworthy that inequality in wealth conservative economies is lower with wealth-based tax regardless of $\lambda$ for both universal and optimal targeted redistributions. Taking the case when the tax rate is similar to the OECD countries apply ($\lambda \approx 1/3$) and targeted redistribution as an example, one can see that $G_1$ is approximately 4.5 times greater than $G_0$. Note that a totally egalitarian society ($G = 0$) is not achieved with tax on trade even with $\lambda = 1$, which is unrealistic but a possible result when $\epsilon = 0$ and $\tau = 1$.

\begin{figure}[!htbp]
    \centering
    \includegraphics{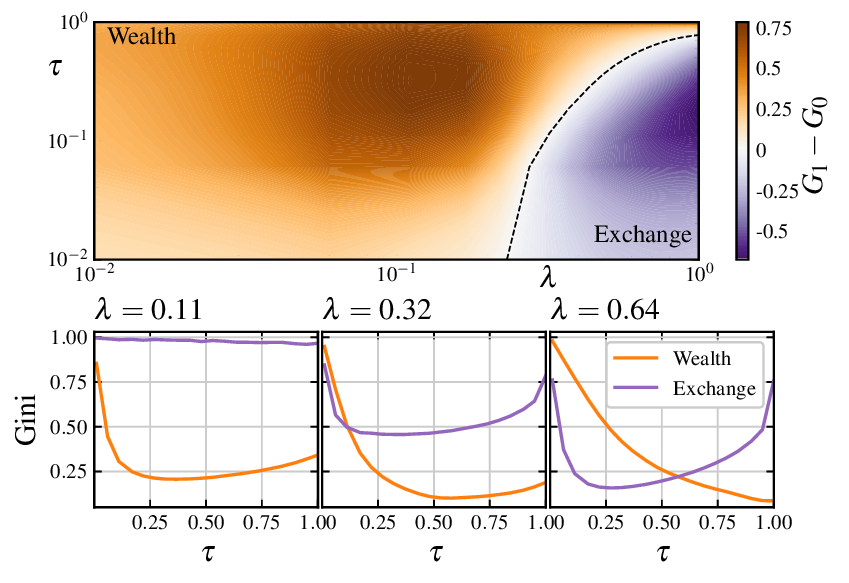}
    \caption{Top: Difference between equilibrium $G_1$ and $G_0$ and its dependency on $\lambda$ and $\tau$. The orange region represents the combinations of tax rates and target fractions where wealth-based tax is more efficient than the transaction-based. The dashed line shows where both policies are equivalent regarding the reduction of inequality. Bottom: Gini cofficients for both taxation systems as functions of the $\tau$ for three different tax rates: $\lambda = 0.11$, 0.32, and 0.64.}
    \label{fig:phase}
\end{figure}

This leads to the question: ``Is there a specific combination of tax rate and target fraction where one system is superior to the other?''  We seek to address this question by illustrating the difference $G_1-G_0$ for various pairs $(\lambda, \tau)$ in the top panel of figure~\ref{fig:phase}. Positive values of $G_1 - G_0$ indicate that wealth-based taxation is more effective, whereas negative values mean that trade taxation is preferable for diminishing inequality. One can see that for taxes larger than 0.2 the trade taxation has greater impact depending on the target. 

The lower panels displayed in Fig.~\ref{fig:phase} illustrate the relationship of $G$ and $\tau$ across three distinct tax rates. 
When $\lambda = 0.11$, clearly, there is no advantage of applying a transaction-base tax system as the Gini index stays almost at the maximum value no matter the fraction of target receivers ($\tau$) is, while for that case, the Gini index changes dramatically in the wealth based system with Gini being lower than 0.25 if $\tau \approx 0.25$.
The other two plots ($\lambda = 0.32$ and $0.64$) indicate that exchange-based taxation is preferable, for low population redistribution fraction: $\tau < 0.1$ in the middle panel, $\tau < 0.55$ in the right one.

\subsection{Wealth and transaction taxes}
\begin{figure}[!htbp]
    \centering
    \includegraphics{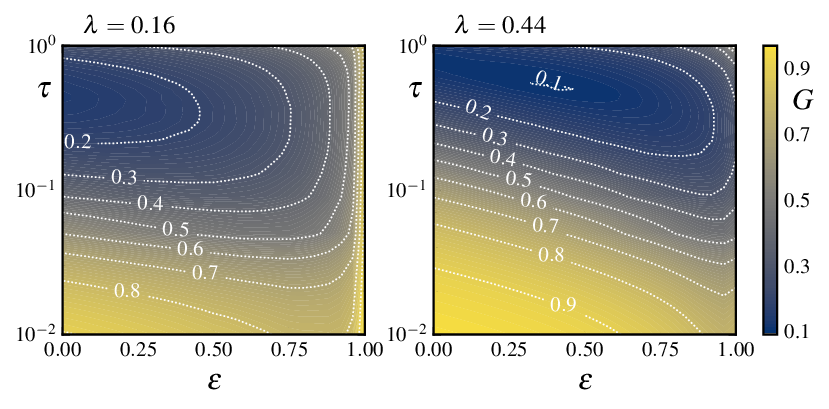}
    \caption{Impact of the target $\tau$ and the tax combination $\epsilon$ parameters on the Gini coefficient for two tax rates ($\lambda = 0.16$ and 0.44) in combined taxation systems.}
    \label{fig:contour-eps}
\end{figure}

The taxation system in real economies takes into account a combination of taxes on wealth and transactions. This way, we considered mixtures of both and how it impact on the inequality. We vary $\epsilon$ from 0 to 1 for each target and tax rates. Interestingly, we find that for $\lambda \geq 0.39$ there are combinations of $\lambda$ and $\tau$ in which this hybrid taxation is preferable. Representative $G(\epsilon, \tau)$ for two different tax rates are shown in Fig.~\ref{fig:contour-eps}. In the left panel ($\lambda = 0.16$) one observes that the minimum Gini coefficient, $G_\text{min}$, is approximately 0.14, which occurs when solely wealth is taxed. Conversely, in the case when $\lambda = 0.44$ (right panel), $G_\text{min} \approx 0.099$ when $\epsilon$ remains between 1/3 and 0.48. This reduction in $G_\text{min}$ is an indication that mixed taxation policies may be the best option to reduce inequality. In both cases, the target for achieving the minima should be larger than 50\% of the poorest agents. Most of the governmental projects, however, restrict the allocations to a very small portion of the poorest individuals, resulting in negligible efficiency in terms of Gini coefficient reduction.  For a significant impact on reducing inequality, a broader target is necessary in both wealth and trade bases, as well as in a combination of the two. This ensures that the effects of the taxation approach results into real and substantial reductions in inequality.

\section{Conclusion}

The trend of increasing economic inequality leads to a state of condensation, where the economy reaches its ``thermal death'', characterized by the absence of wealth flowing among individuals. This situation highlights the necessity for state intervention. Taxation and subsequent redistribution are perhaps the most common methods to impede the rising inequality. Through agent-based simulations we examine the impact of trade and wealth bases on economic inequalty in wealth-conservative economies.

Our results demonstrate the existence of an optimal redistributional target, corresponding to the minimum $G$ for each tax rate, in both taxation systems. Notably, wealth-based taxation outperforms taxation on exchange in universal and optimal targeted redistributions. For non-optimal targets we see a phase diagram that exhibits two different regions, each one related to the higher impact on lowering $G$ for the respective taxation system.

Considering combinations of wealth and trade tax bases, our results show that mixtures of the taxation systems can have greater impact in reducing the Gini coefficient. Specifically, we notice that for $\lambda \geq 0.39$ collecting approximately 1/3 of total tax rate on the exchanges is necessary to achieve the $G_\text{min}$. On the other cases taxation solely on wealth is preferable in the context of minimizing the inequality.

In spite its higher effectiveness in reducing inequality, wealth-based taxation contributes a smaller share to the total tax revenue compared to trade-based taxation, which encompasses consumption, income, and various forms of wealth transfers. Our findings underscore the importance of striking a balance between the two taxation systems to achieve effective reduction of economic inequality.

\bibliographystyle{unsrtnat}
\bibliography{references.bib}
\end{document}